\documentclass[12pt]{article}
\usepackage{latexsym,epsfig,graphicx,amsmath,amssymb,amscd,undertilde,multirow,chicago,psfrag,paralist,dsfont,subcaption,float}
\usepackage[titletoc]{appendix}
\newcommand{\thetavec}{{\boldsymbol{\theta}}}

\newcommand{\Sigmavec}{{\boldsymbol{\Sigma}}}

\newcommand{\zerovec}{{\boldsymbol{0}}}

\newcommand{\N}{{\textrm{N}}}

\begin{document}

%%%%%%%%%%%%TITLE%%%%%%%%%%%%%%%%%%%%%%%%%%%%%%%%%%%
\title{Reliability Analysis of Polymeric Materials}
%%%%%%%%%%%%%%%%%%%%%%%%%%%%%%%%%%%%%%%%%%%%%%%%%%%%%%%%%%%%%%%%%%%%%%%%%%%%%%%%%%%%%%%%%%%%%%%%%%%%%%%%%%%%%%%%%

\author{Caleb King\\
Statistical Sciences Group\\
Sandia National Laboratories\\
Albuquerque, NM 87185\\
\and
Zhibing Xu\\
Quantitative Analytics \& Modeling\\
Freddie Mac\\
McLean, VA 22102\\
\and
I-Chen Lee\\
Department of Statistics\\
National Cheng Kung University\\
Tainan, Taiwan 701\\
\and
Yili Hong\\
Department of Statistics\\
Virginia Tech\\
Blacksburg, VA 24061\\
}

\date{\today}

\maketitle
%%%%%%%%%%%%%%%%%%%%%%%%%%%%%%%%%%%%%%%%%%%%%%%%%%%%%%%%%%%%%%%%%%%%%%%%%%%%%%%%%%%%%%%%%%%%%%%%%%%%%%%%%%%%%%%%
\begin{abstract}
Polymeric materials are widely used in many applications and are especially useful when combined with other polymers to make polymer composites. The appealing features of these materials come from their having comparable levels of strength and endurance to what one would find in metal alloys while being more lightweight and economical. However, these materials are still susceptible to degradation over time and so it is of great importance to manufacturers to assess their product's lifetime. Because these materials are meant to last over a span of several years or even decades, accelerated testing is often the method of choice in assessing product lifetimes in a more feasible time frame. In this article, a brief introduction is given to the methods of accelerated testing and analysis used with polymer materials. Special attention is given to degradation testing and modeling due to the growing popularity of these techniques along with a brief discussion of fatigue testing. References are provided for further reading in each of these areas.


\end{abstract}

%%%%%%%%%%%%%%%%%%%%%%%%%%%%%%%%%%%%%%%%%%%%%%%%%%%%%%%%%%%%%%%%%%%%%%%%%%%
\section{Introduction}
%%%%%%%%%%%%%%%%%%%%%%%%%%%%%%%%%%%%%%%%%%%%%%%%%%%%%%%%%%%%%%%%%%%%%%%%%%%
Polymeric materials are widely used in many applications such as paints, coatings, and insulations. For example, photovoltaics (PV) systems contain many polymeric components, such as encapsulants, frontsheets, backsheets, edge sealants, and junction boxes. Other polymeric materials include thermoplastic, thermosetting, and elastomeric materials. Polymer composites, which are made from either combinations of polymers or combinations of polymers with other materials, are an important class of materials. They have many desirable properties, such as comparable levels of strength and endurance with metal alloys, while also remaining lightweight and economical. As such, polymer composites have many applications in transportation, manufacturing, and alternative energy production industries. It is for this reason that the reliability of polymeric materials and polymer composites is an important area for research and development.

Despite their various desirable traits, polymeric materials in the field are still susceptible to degradation and, over extended periods of time, they tend to lose their functional properties (such as tensile strength) under harsh environmental conditions. Because the timescale in which this degradation occurs can be on the order of several years or even decades, accelerated degradation testing (ADT) is often used to evaluate long-term performance of such polymeric materials. There are two primary types of ADT, which are determined by the manner in which the data are collected. If the measurement of degradation involves destructive testing (see \textbf{stat02181}), then the test is known as accelerated destructive degradation testing (ADDT). For ADDT data, only one measure is available per sample unit and so tends to require more samples. However, if the degradation measurement is nondestructive, then multiple measures are available per sample unit and so this form of testing is known as repeated measures degradation testing (RMDT).  For polymer composites, one key objective is to demonstrate the material has sufficient life under fatigue. Accelerated life test (ALT) methods are usually used to collect data and make inference about the fatigue lifetime of the material in the field (see \textbf{stat02159} for more information about life testing methods).

In this article, we provide introductory level information on the reliability analysis of polymeric materials and polymer composites. Section~2 focuses on RMDT data modeling and analysis for polymeric materials. Section~3 focuses on ADDT data modeling and analysis. Section~4 focuses on ALT data analysis and test planning for fatigue testing of polymer composites. Section~\ref{sec:conclusion} contains some concluding remarks.

\section{Repeated Measures Degradation Testing}

\subsection{Definition}

\begin{figure}%[H]
\begin{center}
\includegraphics[width=0.6\textwidth]{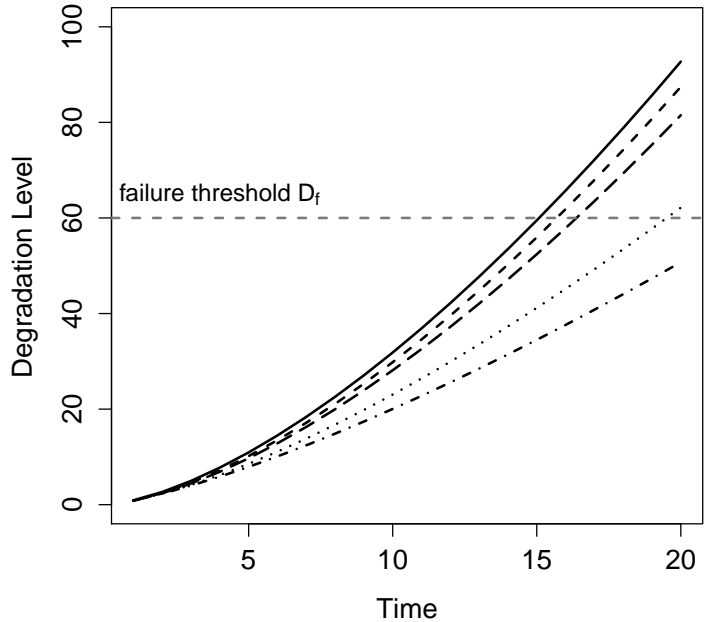} \\
\caption{Illustration of degradation paths for several product units.}\label{fig:rmdt}
\end{center}
\end{figure}

For products with long lifetimes, but showing gradual reduction of performance or reliability assets, the repeated measures degradation test (RMDT) provides a more informative form of testing than traditional life testing. Because it involves non-destructive measures, RMDT has the ability to still assess lifetime through performance degradation while still allowing for the possibility of retaining usable life. Once a performance measure of degradation has been specified, the lifetime can be determined when the degradation measure has cross some specified threshold ($D_f$).  For example, if the luminosity of a light-emitting diode (LED) decreases to a certain level, such as 60\% of the original luminosity, the LED is defined as a failure. This is known in the literature as a ``soft failure" as opposed to the more commonly known ``hard" or ``catastrophic failure". Figure~\ref{fig:rmdt} shows the repeated degradation measurements of five product sample units with $D_f$ at 60.

The two most commonly used models for analyzing RMDT data are the general degradation path (GDP) models, which assume a specified functional form for the degradation path of a sample, and stochastic process models, which model the degradation process as a stochastic process such as the Wiener process (e.g., \citeNP{Whitmore1995}; see \textbf{stat03025}). \citeN{LuMeeker1993} first introduced the use of GDP models to fit degradation trends and gave the failure-time prediction based on Monte Carlo simulation (see \textbf{stat06174.pub2}). Since then, a growing area of research in the analysis of RMDT data have utilized the GDP model, examples including \citeN{MeekerEscobarLu1998}, \citeN{BaeKvam2004}, \citeN{PechtFan2012}, and \shortciteN{HongDuanetal2015}. Examples of stochastic process models include \citeN{BagdonaviciusNikulin2001}, \citeN{LawlessCrowder2004}, \citeN{WangXu2010}, and \citeN{YeChen2014}. Reviews of literature in modeling RMDT data can be found in \citeN{MeekerHongEscobar2011}, and \citeN{YeXie2015}.

\subsection{Methodology}

\subsubsection{General Degradation Path Model}
Suppose there are $n$ samples with degradation measurements taken at times $\{t_{i1}, t_{i2}, \cdots, t_{ik_i} \}$, where $i=1, \cdots, n$ and $k_i$ is the number of times at which sample $i$ is measured, which need not be the same for all samples. The degradation measurement $y_i(t_{ij})$ is then modeled by a linear or nonlinear function:
\begin{align*}
y_i(t_{ij})=f(t_{ij}; \thetavec, \boldsymbol{\varphi}_i) + \epsilon_{ij}, \quad  i= 1, \cdots, n,\ j=1, \cdots, n_{k_i},
\end{align*}
where $\thetavec$ is a vector of fixed effects parameters, and $\boldsymbol{\varphi}_i$ is a vector of random effects, which are used to describe the unit-to-unit variability around the population trend. The random effects are usually assumed to follow a multivariate normal distribution $\N(\zerovec, \Sigmavec)$ and the measurement error term $\epsilon_{ij}$ is assumed to follow a normal distribution $\N(0, \sigma^2)$. The general path function may be derived from physical or chemical mechanisms or be specified empirically. \citeN{MeekerHongEscobar2011} provides some examples of functional forms. To describe the effect of the acceleration variable (e.g., temperature), a component of the fixed effect parameters $\thetavec$ are linked to the acceleration variable through a functional relationship. For example, in temperature acceleration, the Arrhenius function is the common model of choice to link the temperature to the degradation path (see \textbf{stat02199}). A good reference for acceleration variables is \citeN{EscobarMeeker2006}.

In most cases, the acceleration factors are constant over time. While this is mostly true for lab testing, in the field it is more common for the acceleration factors to vary with time, such as the daily temperature and use rate. Thus, recent research has focused on the inclusion of time-varying or dynamic covariates into the general path model, usually through accumulating covariate effect functions (e.g., cumulative damage). Recent work on modeling time-varying covariates on degradation of polymeric materials include \shortciteN{HongDuanetal2015} and \shortciteN{XuHongJin2015}, in which both linear and nonlinear functions are used to incorporate outdoor weather variables (i.e., ultraviolet exposure, temperature and relative humidity).

To predict material reliability, the time to failure distribution needs to be estimated. With a threshold $D_f$ defined, setting $D_f=f(t_f; \thetavec, \boldsymbol{\varphi})$ and solving for $t_f$ allows for prediction of failure times. The distribution of $t_f$ will then give the lifetime distribution. If $f(\cdot)$ is a linear function of the random effects, it is possible to derive a closed form for this prediction. In general, this is not often the case and so the predictions are often found using Monte Carlo simulation. \citeN{MeekerHongEscobar2011} provide examples for lifetime estimation based on RMDT models.

\subsubsection{Stochastic Process Models}
The stochastic process is often used when assessing the amount of erosion, crack, and creep in a material. Among the many degradation models available, the Wiener process model and gamma process model are the two most popular, though the inverse Gaussian model also appears quite often in literature. Using the gamma process as an example, the degradation measurements $y(t)$ are modeled by a gamma process characterized by a trend function $\eta(t)$ and a scale parameter $\xi$. The increment $\Delta y(t)=y(t+\Delta t)-y(t) $ follows a gamma distribution $Gamma\{\xi, \eta(t+\Delta t)-\eta(t)\}$. The corresponding probability density function (pdf) of $\Delta y(t)$ is
\begin{align*}
f[\Delta y(t); \Delta\eta(t), \xi]=\frac{1}{\Gamma[\Delta\eta(t)] \xi^{\Delta\eta(t)}} \Delta y(t)^{\Delta\eta(t)-1} \exp\left[-\frac{\Delta y(t)}{\xi}\right],
\end{align*}
where $\Gamma(\cdot)$ is the gamma function, $\eta(t)$ is a non-decreasing function, and $\Delta \eta(t)=\eta(t+\Delta t)-\eta(t)$.

Under acceleration, the trend function can be related to the acceleration factor by using a functional relationship. Examples of functional relationships include the Arrhenius, power law and the exponential relationships. \citeN{MeekerHongEscobar2011} provide examples for stochastic process modeling of RMDT data. Further reference in stochastic process model includes \citeN{BagdonaviciusNikulin2001}, \citeN{LawlessCrowder2004}, \citeN{YeChen2014}, and \citeN{HeonsangLim2015}.

\section{Accelerated Destructive Degradation Testing}

\subsection{Definition}

While non-destructive tests can yield invaluable information while leaving the product relatively undamaged, there are situations in which key material properties can only be assessed through destructive means. A prime example is in trying to assess material strength, which often can only be done by measuring the force at which the material breaks. When destructive testing is utilized in accelerated degradation testing, the resulting testing procedure is referred to in the literature as accelerated destructive degradation testing (ADDT).

\subsection{Methodology}
Because of the destructive nature of the testing, there are slight differences between the construction, implementation, and analysis of an ADDT and an RMDT. One key difference is the fact that an ADDT requires many more samples than for an RMDT as each unit under test is permanently removed from the experiment after measuring. The samples are usually grouped into independent ``batches'' with measurements from the same batch tending to be correlated with one another (see \textbf{stat02430}).

The general degradation path model from RMDT can be adapted here with a slight alteration. The degradation measurement is now modeled as
\begin{align*}
y_i(t_{ij})=f(t_{ij}; \thetavec, \rho) + \epsilon_{ij}, \quad  i= 1, \cdots, n,\ j=1, \cdots, n_i,
\end{align*}
where $\rho$ is the within-batch correlation and $j$ now represents the $j$th unit in batch $i$. It is often assumed that the correlation is the same for each batch, although this can easily be generalized.

For modeling and analysis of an ADDT for polymer materials, there are three primary methods available: the traditional approach used in industry and the parametric and semi-parametric methods that have been proposed in the statistics literature. An R package has been developed for use by industrial statisticians that implements each of these methods (see \shortciteNP{XieJinHongVanMullekom2017}, and \shortciteNP{JinXieHongVanMullekom2017}).

\subsubsection{The Traditional Method}

In the traditional method specified in the industry standards \citeN{UL746B}, $f(t_{ij}; \thetavec, \rho)$ is  estimated in two steps. In the initial step, a cubic polynomial regression model is fit to the response as a function of time and used to estimate the time to failure. It is important that data exists beyond the threshold as polynomial regression cannot accurately extrapolate beyond the observed data. In the next step, a second regression is performed on the estimated times as a function of the acceleration factor (often using a linear regression model). While the traditional method has the advantage of making no assumptions regarding the degradation model form and is simple to implement using basic statistical procedures, there is no explicit method for assessing uncertainty in the final result. In addition, there is no consideration of the correlation parameter $\rho$ in the method. More information about this procedure can be found in \citeN{UL746B}  and \shortciteN{Kingetal2017}.

\subsubsection{The Parametric Method}

In the parametric method, a specific form for $f(t_{ij}; \thetavec, \rho)$ is specified that describes the relationship between time, acceleration factor, and the response in a single concise form. This model is then fit to all of the data at one time, usually via the maximum likelihood procedure (see \textbf{stat04464}) with $\boldsymbol{\epsilon} \sim \text{Normal}(\boldsymbol{0},\boldsymbol{\Sigma})$, where $\boldsymbol{\Sigma}$ is a correlation matrix. The primary advantage of this method is the ability to quantify uncertainty with relative ease and allows for prediction of the response beyond the observed data, which can be useful in experimental planning and implementation. However, this method might be sensitive to the choice of model and so may yield biased results should the model be misspecified. More information about this procedure can be found in \shortciteN{Kingetal2017}.

\subsubsection{The Semi-parametric Method}

In the semi-parametric method, $f(t_{ij}; \thetavec, \rho)$ is specified using a nonparametric method to model the relationship between the response and time and a parametric model to relate the nonparametric portion to the acceleration factor (see \textbf{stat05781}). In order to preserve the monotonic nature of degradation, monotonic splines, such as B-splines, are the primary tool of choice. Just as with the fully parametric approach, the model can be fit to the data all at one time. The primary advantage of this method is the reduction of the risk of model misspecification in the degradation behavior. However, while it is still possible to extrapolate on the acceleration factor, this method does not necessarily allow for extrapolation in the time domain, which may cause issues if all observations do not reach the failure threshold. More information about this procedure can be found in \shortciteN{Xieetal2016}.

\subsubsection{Thermal Indexing}

\begin{figure}%[t!]
\centering
\begin{subfigure}[t]{0.5\textwidth}
\centering
\includegraphics[scale=0.5]{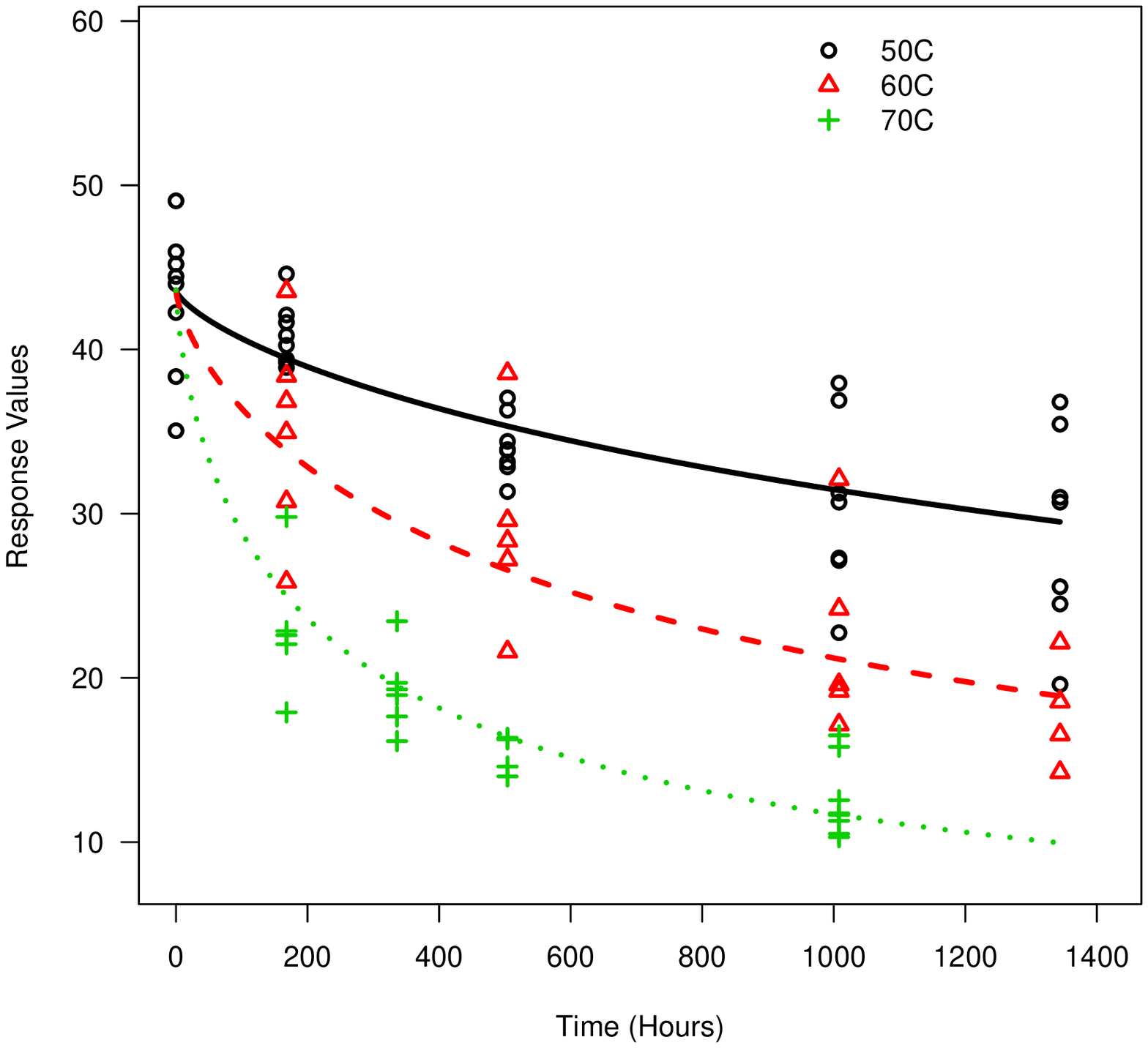}
\caption{ADDT Data and Model Fit}
\end{subfigure}%
\begin{subfigure}[t]{0.5\textwidth}
\centering
\includegraphics[scale=0.45]{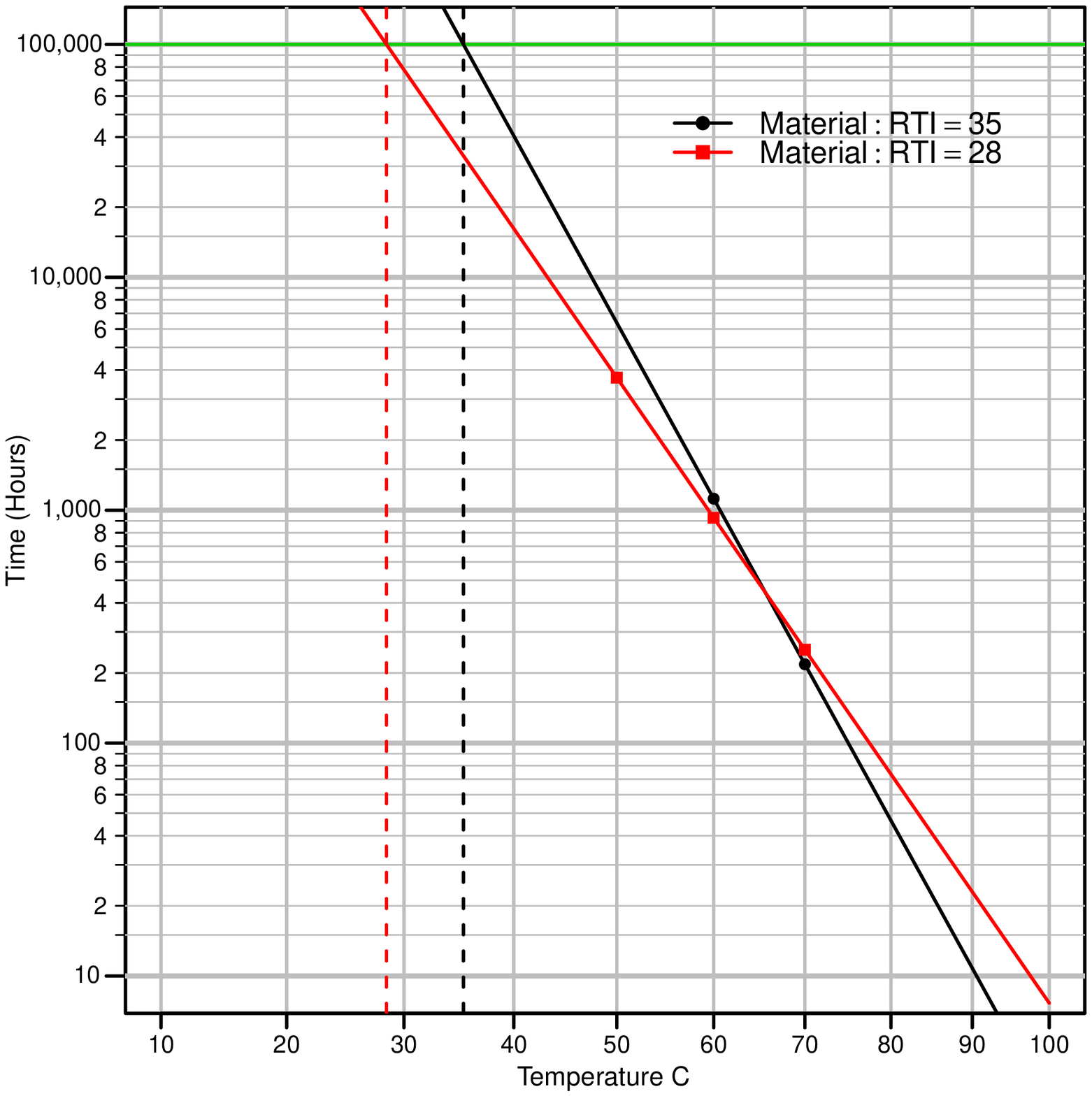}
\caption{Thermal Index Plot}
\end{subfigure}
\caption{Example of thermal indexing procedure.}\label{fig:thermal.index}
\end{figure}

One common area of application of ADDT with polymer materials is known as \emph{thermal indexing}. Here, the interest is on the thermal effects on a particular characteristic of a polymer material, most commonly its strength. Assuming an Arrhenius model of strength as a function of inverse Kelvin temperature, an ADDT is performed to assess the specific parameters of the relationship. Then, the model is used to extrapolate to a defined lifetime (e.g. 100{,}000 hours) and then extract the temperature at which that lifetime is achieved. This temperature is called the thermal index. Thermal indexing is primarily used for comparing materials to known materials from a database or to material created by a competitor. In the latter case, a separate relationship for the competitor material is estimated and then used to compute a relative thermal index. More details about this application can be found in \shortciteN{Kingetal2017}. An example of the procedure can be found in Figure~\ref{fig:thermal.index}.

\section{Fatigue Testing}

\subsection{Description}

Fatigue occurs when a material is exposed to varying levels of stress over a period of time. This makes fatigue testing an important part of assessing the reliability of the material. Due to the long-term durability of polymer composite materials, accelerated life testing (ALT) methods are often utilized to collect failure information in a shorter period of time.

\begin{figure}%[t!]
\centering
\begin{subfigure}[t]{0.5\textwidth}
\centering
\includegraphics[scale=0.5]{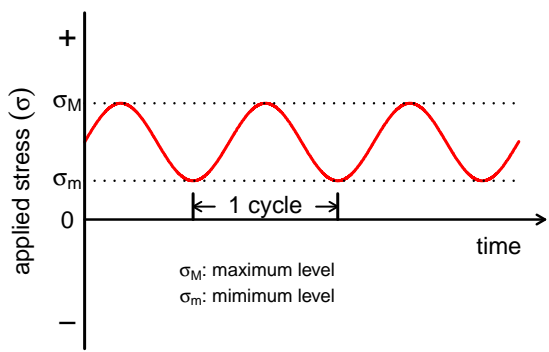}
\caption{Example of Constant Amplitude Cyclic Fatigue Testing} \label{fig:con_loading}
\end{subfigure}%
\begin{subfigure}[t]{0.5\textwidth}
\centering
\includegraphics[scale=0.25]{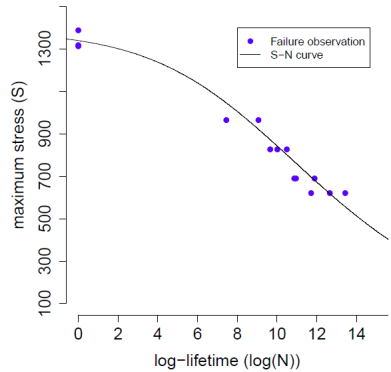}
\caption{Fatigue Data and Fitted S-N Curve} \label{fig:data_SN}
\end{subfigure}
\caption{Example of fatigue testing and data.}\label{fig:ALT}
\end{figure}

Much of the testing performed in this field is in accordance with the standards provided in \citeN{Standard10e739}. The most commonly used method is constant amplitude cyclic fatigue testing, in which a tensile or compressive stress is applied to a test coupon and cycled through a specified maximum and minimum level. The lifetime is measured by the number of cycles until catastrophic failure. Figure~(\ref{fig:con_loading}) is a graphical example of the constant amplitude cyclic fatigue testing. The maximum stress is usually defined as the acceleration factor.

Much of the statistical research in this area consists of defining failure models (see \textbf{stat02141.pub2}) and, in some cases, deriving optimal test plans based on a particular model. There are a wide variety of models that have been developed for modeling fatigue data (e.g., \citeNP{Wohler1870}, \citeNP{Weibull1949}, \citeNP{Spindel1981}, \citeNP{Kohout2001}, and \citeNP{Epaarachchi2003}), most of which relate the log-lifetime to the maximum stress through a relationship known in the field as the S-N curve.  Figure~(\ref{fig:data_SN}) shows an example of the fatigue data and its corresponding fitted S-N curve. In terms of test planning, the standard procedures tend to use balanced designs. However, given an appropriate fatigue model, it is possible to derive a test plan that is optimal for that model. Much of the research in the statistics literature is focused on developing these optimal plans (see \textbf{stat04090.pub2}).

\subsection{Methodology}

\subsubsection{Balanced Designs}

In balanced test plans, the levels of maximum stress are equally spaced across the range of stresses being tested and the total sample size is equally distributed among the test levels. While this has the advantage of being simple in layout and easy to implement, it is not necessarily the optimal test plan configuration. In general, it is better to allocate more samples to the lower stress levels as they lie closer to the levels of interest, a technique often associated with statistically optimal designs.

\subsubsection{Statistically Optimal Designs}

In the statistics literature, the lifetime $T$ in cycles to failure is treated as a random variable. Often the distribution of this random variable is taken to be a member of the log-location- scale family (see \textbf{stat05887}), with cumulative distribution function (cdf) and pdf given as
\begin{equation*}
F\left(t\right)=\Phi\left[\frac{\log\left(t\right)-\mu}{\nu}\right], \quad \textrm{ and }\quad
f\left(t\right)=\frac{1}{\nu}\phi\left[\frac{\log\left(t\right)-\mu}{\nu}\right],
\end{equation*}
respectively, where $\mu$ is a location parameter and $\nu$ is the scale parameter. Here, $\Phi(\cdot)$ and $\phi(\cdot)$ are the cdf and pdf of the standard distribution with $\mu=0$ and $\nu=1$. The most common examples of the log-location-scale family are the Weibull and lognormal distributions (see \textbf{stat07358} and \textbf{stat05889}, respectively).

In fatigue modeling, it is usually assumed that the scale parameter is constant and the location parameter is a function of the stress. Once this model has been specified, key quantities of the lifetime distribution at the normal use condition can then be estimated. An optimal test plan will generally allocate resources that minimize the variance of this estimator.

\noindent\emph{Traditional Optimum Design}

Traditional methods for designing optimal fatigue tests are based on properties of maximum likelihood estimators. As such, their goal is to minimize the asymptotic variance of an estimator based on the Fisher information matrix. Examples of this method can be found in \citeN{meekerescobar1998} and \shortciteN{Kingetal2016}. Such optimal designs may depend on knowing the parameter values, referred to as planning values in the literature, making it necessary to assess the sensitivity of a proposed plan to these planning values.

\noindent\emph{Bayesian Design}

When model parameters must be known prior to testing, the test planning easily lends itself to Bayesian techniques (see \textbf{stat00207.pub2}). Such designs can incorporate prior uncertainty regarding the model parameters into the test plan and then derive optimal test plans based on the expected posterior uncertainty over multiple test results, known as the pre-posterior expectation. Prior information can be obtained by either historical data or subject matter expertise. More details regarding Bayesian test planning can be found in  \citeN{ZhangMeeker2005}, \citeN{ZhangMeeker2006}, and \shortciteN{Hongetal2015b}.\\

\noindent\emph{Sequential Bayesian Design}

Even under accelerated conditions, testing can still last several weeks or even months. In addition, most testing laboratories are sometimes equipped with only one or two testing machines, which makes testing multiple samples in parallel almost impossible. In these conditions, a sequential testing procedure can prove invaluable (see \textbf{stat04107}). Often utilizing Bayesian methods, after initial design points have been tested, the next optimal design point is determined by minimizing the asymptotic variance of a prediction over the posterior distribution of parameters given the current data. This process is then repeated over the course of the test until all samples have been used or the desired variance level has been reached. More information about sequential Bayesian design can be found in \shortciteN{Leeetal2017}.

\section{Concluding Remarks}\label{sec:conclusion}
In this article, we provide an introductory level description for statistical methods that can be used for reliability analysis of polymeric material. We focus on accelerated testing and analysis for the reliability study of polymer materials. In particular, we cover accelerated degradation testing, accelerated destructive degradation testing, and fatigue testing, because these methods are widely used in the reliability analysis of polymeric materials and composites. While it is challenging to provide a coverage of all the topics in greater depth in one short article, references have been provided for each topic so that readers can have the relevant sources to find further information.

\section{Acknowledgments}
The authors would like to thank the Editor for his helpful comments and suggestions on an earlier version of this article. The authors acknowledge Advanced Research Computing at Virginia Tech for providing computational resources. The work by Hong was partially supported by the National Science Foundation under Grant CNS-1565314 to Virginia Tech.

\end{document}